# A Management Approach of an E-Tutoring Program for High School Students

Spyridon Doukakis

Department of Informatics, Ionian University, Greece Pierce-The American College of Greece, Greece

*ABSTRACT*

*The inclusion of e-tutoring programs to support secondary school students is an international practice that is reinforced by both the education policies of the Ministries of Education and the potential of technology. The operation and management of the relevant programs is a challenging process, as the goal is to effectively support students and improve their learning. In the present work, the management approach of an e-tutoring program that operates from the school year 2012-2013 is presented. The approach includes a) the presentation of the processes through which the e-tutoring program is carried out, b) the information systems for monitoring the progress of its operation, such as students' participation, the duration of their participation, their learning needs, the correlation with their performance and c) the e-tutors' training procedures. In addition, practices are emerging that offer a comprehensive monitoring framework, which favors the proper functioning and further development, as well as the increasing participation of students.*

## 1. Introduction

The growing online communication capabilities have identified new opportunities in education. Researchers and educators study and highlight appropriate ways to use technologies to support student learning. One of the educational sectors that has been significantly affected by these developments is distance education. Educators and learners, through appropriate environments, have the opportunity to collaborate in real time by maintaining continuous interaction and using speech and text. Today, the idea that through the Internet all forms of education can be made without boundaries, it may be really happen.

The implementation of distance education with the use of digital technologies has also affected the field of school education. Virtual schools have been developed, which offer exclusive distance learning and courses, through appropriate environments [1, 2]. At the same time, the schools that operate in building facilities, have strengthened their structures with synchronous and asynchronous means of communication. The implementation of distance education emerged as an extremely useful solution during the COVID-19 period where thanks to digital tools, all education units were enabled to continue to offer learning opportunities.

E-tutoring programs run in real time and are used by students to support and enhance their learning. The purpose of this study is to highlight a comprehensive e-tutoring program management approach and to identify critical milestones that can contribute to the success of such programs. In the following paragraphs, the principles of the e-tutoring framework will be introduced. Afterwards, a case of an e-tutoring program will be described along with the program

    

management framework. This study will be completed with some proposals in order schools and educational systems around the world, to be able to include e-tutoring programs in their educational process.

## 2. LITERATURE REVIEW

E-tutoring is a distance learning service that utilizes digital technologies. It has similar characteristics to the characteristics of teaching in a classroom. It is a teaching model, where the teacher facilitates students to acquire knowledge, develop skills and change attitudes towards the subject [3] and at the same time, one or more students generally seek access to knowledge or seek support / need clarification about a specific learning need. Differences are observed in the student-teacher collaboration environment. Thus, through a customized online environment that includes various support tools, e-tutoring is implemented [4].

Although e-tutoring and e-learning have common features there are significant differences. Initially, students are not required to participate in e-tutoring, nor are they required to prepare specific work to participate in e-tutoring. Instead, students connect to the online e-tutoring environment additionally to their daily school program. The goal of the students is to receive support to overcome obstacles they encountered while studying the material of their regular school program. For example, they can use e-tutoring program, if they did not understand a concept and need some clarification, if they want extra support for an issue, or if they need some help to solve an exercise or complete a task. In e-learning, students have clear tasks, deadlines, mandatory participation, etc., elements that do not exist in e-tutoring [3]. Therefore, the e-tutoring takes place after the completion of the compulsory school program, in order to further support the students.

In different countries, e-tutoring is offered by public, private and non-profit organizations. The implementation of e-tutoring programs in the United States is constantly increasing due to the low cost of the service [5]. In 2020 the increase in the US is about 8% with e-tutoring emerging as the most popular form of private education [6]. E-tutoring programs have been implemented with great success in Canada, Korea and Taiwan.

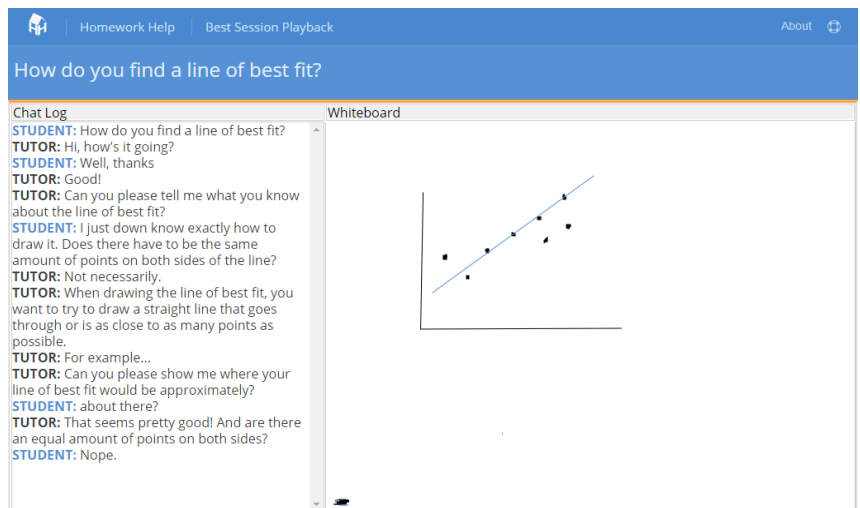

Figure 1. Snapshot from an e-tutoring meeting at Homework Help

The Ontario e-tutoring program was started in 2008 by the Canadian Ministry of Education and is funded exclusively by the Ministry (https://courses.elearningontario.ca/). It is offered free of charge to Ontario secondary students, five days a week, from Sunday to Thursday, from 5:30 p.m. until 9:30 p.m., giving to students extra learning opportunities. The teachers have been trained and certified by the Canadian Ministry of Education. Using a secure digital environment, students in grades 7 to 10 with certified educators work together, in order the latter to support students in Mathematics through synchronous communication. In this way, more than 20,000 students are supported confidentially and anonymously. Figure 1 shows a snapshot of an e-tutoring meeting at Home Help.

The Korean government in 2004 designed and developed the Cyber Home Learning System (CHLS). The government's goal was to improve student performance and reduce education costs for families. For this purpose, a specific legislative framework was developed. CHLS operates in all sixteen (16) regional education offices in Korea. The number of teachers is adjusted based on participation, which is increasing, as more and more families are taking advantage of the dynamic aspects of the program. Figure 2 shows the basic services of CHLS (Learning management, study management, guidance and counseling, self-teaching, diagnostic assessments and teleconferencing). The services are provided to primary school students through the Korean Ministry of Education, while to high school students through the Korean Institute for Educational Development [7].

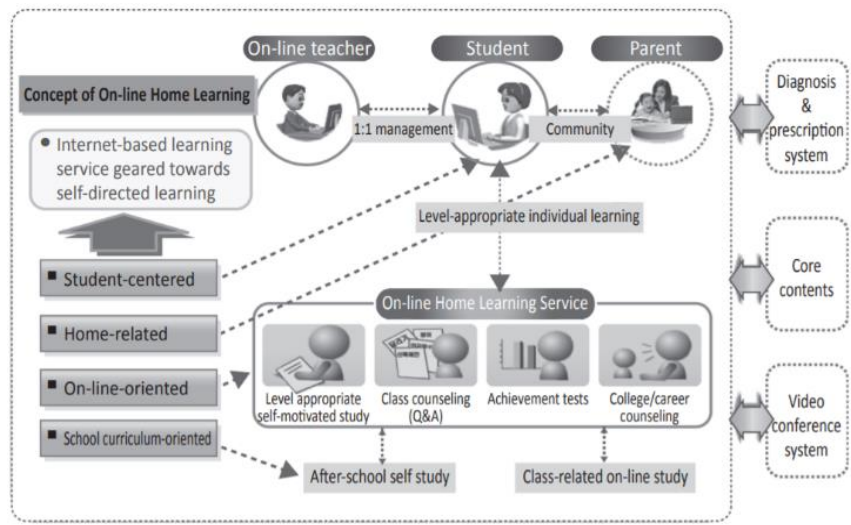

Figure 2. Conceptual structure of Cyber Home Learning System, [8]

With the initial goal of supporting families living far from urban areas, in 2006 the Taiwanese government launched the Online Tutoring for After School Learning program [9]. In this program, through an e-learning platform, university students provide support to students in real time (see Figure 3). Through the program, students received support and improved their performance.

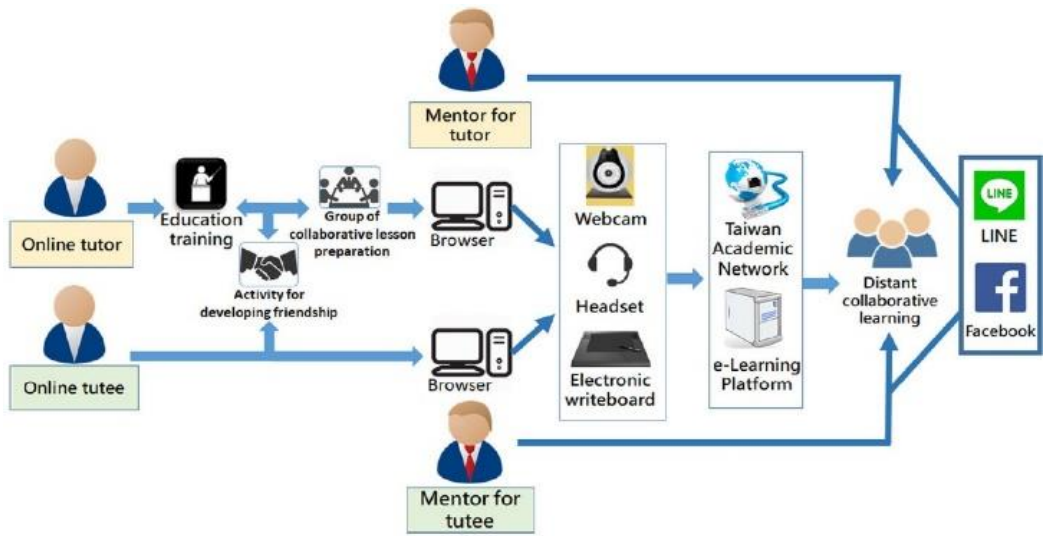

Figure 3. Framework of the Taiwan e-tutoring system

Several studies have been published in recent years on e-tutoring. Some of these studies provide empirical data for programs in the United States, England, Canada, Spain, Scotland, Mexico, Australia and Greece [10]. Studies a) highlight the potential benefits of implementing e-tutoring programs and b) show the improvement of student performance [10]. Researchers point out that e-tutoring under certain conditions can be more effective than traditional classroom teaching if a) the frequency of interaction is high, b) there is immediate feedback and c) the e-tutor focuses on the needs of the students that s/he supports.

More specifically, studies show that when students participate in an e-tutoring program: a) there is a growing involvement in the learning process, b) there is an opportunity to highlight their interests, c) they have the opportunity to overcome their insecurity of being "exposed" to a question, as may be happens in the conventional classroom [10], d) they have the opportunity to modify their view on learning and their view about a discipline [11], e) their self-confidence is enhanced due to immediate response of e-tutors, f) e-tutors and students are given the opportunity to use pedagogical tools that they cannot use in the traditional classroom [12], g) the idea of cooperation and working in groups is strengthened, since students participate in a learning community that encourages those who in the traditional teaching and learning environment were isolated, h) "two-way learning opportunities" are observed for teachers and students, i) communication is strengthened as the anonymity of students and teachers is maintained [2], j) offers the opportunity for e-tutors to take more into account their students' needs.

Research has also studied issues related to students' emotional experiences with e-tutoring [13]. Researchers point out that the limitations of the online environments may offer few opportunities for developing positive emotions between students and teachers. For this reason, they identify the special role of the educator, in order to enhance the development of a positive climate in the e-tutoring environments. Another aspect of e-tutoring programs is the issue of teacher-student collaboration. The results of research have shown that the frequent use of questions is an appropriate technique to enhance discussion and student extroversion, so that the educator understands the shortcomings and needs of the learner [14].

Another area that researchers have focused on is the techno-economic characteristics of e-tutoring programs. Research shows that the cost of designing, developing and implementing e-tutoring programs is up to five (5) times lower than the cost of other forms of support [15]. In addition, it seems that due to the structure of e-tutoring programs, students and teachers have more free time, are safer and have energy reserves for other activities. It also seems that e-tutoring programs can support families who cannot afford the high cost of education, as well as families who have difficulty accessing existing educational facilities because they live in remote or inaccessible areas [15].

All the above were examined for the design and development of the Pierce e-tutoring online program that operates in Greece. In the next section, the program will be presented, the management approach of the program will be highlighted, and its perspectives will be analyzed.

## 3. PIERCE E-TUTORING ONLINE PROGRAM

In the context of student support and the inclusion of innovative services, the school year 2012-2013 there was a discussion about the inclusion of an e-tutoring program within the school unit, which would operate after the end of the daily program and would aim to support learning of the students. The initial study focused on three issues: a) the added value that the operation of an e-tutoring program would have, b) the ability of teachers to support an e-tutoring program, and c) the tools that would be required to operate an e-tutoring program (see Figure 4) [16].

The innovation of the service and the opportunities that would be offered to the students and their families, led to a high degree of acceptance of such a program in the school unit. In addition, it was decided the program to be offered to students and their families at no extra cost. At the same time, after an invitation, four teachers stated their intention to be trained and carry out a pilot implementation of the program for three months, in order to obtain data on how it works. In addition, after studying the existing e-learning environments, it was decided to use the Blackboard Collaborate platform, as the digital tool between teachers and students. The next step was teacher training. For this purpose, an expert in the field of distance education and online learning undertook the training of 4 teachers, who would act as multipliers if the program continued to operate. Two of the 4 teachers had experience in distance education and online learning [16]. Upon completion of the teacher training, one of them undertook the coordination and the management of the e-tutoring program.

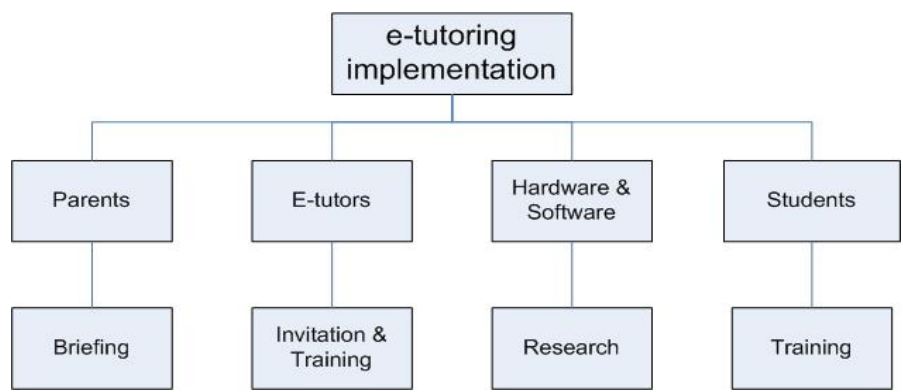

Figure 4. E-tutoring implementation procedure

During the pilot implementation of the program, the Pierce e-tutoring online program was offered in Ancient Greek Language and Mathematics and exclusively for 28 7th grade and 30 8th grade students. The successful pilot implementation led to the full integration of the program for all students from 7th to 10th grade, from Monday to Thursday for 2 hours per day. In the school year 2014-2015, the program was enriched with Physics and Modern Greek Language.

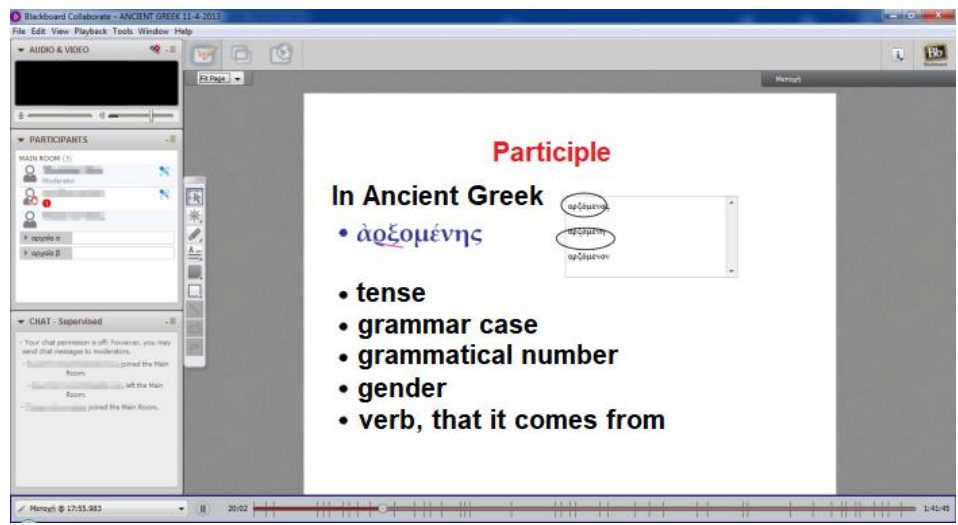

Figure 5. Blackboard Collaborate environment

In order to determine the two hours of operation of the program, the parents were given a questionnaire. From the completion of the questionnaire, the period 19:00-21:00 was selected. As already mentioned, the e-tutor is a teacher from the school unit and is likely to be the teacher of the student who wishes to join the program. The student does not have to stay online for both hours of the program. The length of his/her stay is usually determined by the student himself/herself and his/her personal needs. Thus, a student can connect with one or more (even to the four) programs on the same day.

Many digital tools are used for the operation of the program, such as the sharing of digital material and the development of online activities. The material is mainly developed using presentation, image editing or snapshot tools. The material presented on the whiteboard is fully interactive, as students and educators have the opportunity to intervene by completing, noting, highlighting, deleting, writing directly on the interactive whiteboard using the tools available in the environment (Figure 5).

Also, the possibility provided to the educator to create virtual rooms is a useful tool. By creating groups of students and placing them in virtual rooms, students have the opportunity to collaborate and work on the production of a project. In addition, in the room the student can work individually on his/her specific learning need. Collaborating in virtual rooms is also helpful for students who need to overcome exposure fears in a group. In addition, the ability to assess students with closed-ended questions (True/False and multiple choice) offers opportunities for reflection and self-assessment for students. Another useful tool is chat. Using chat, students and the e-tutor can communicate privately or with the whole team. In addition, emoticons play an important role in communication, facilitating discussion and limiting potential disruption that may occur in a large group [16].

## 4. MANAGEMENT APPROACH OF PIERCE E-TUTORING ONLINE PROGRAM

After the pilot implementation of the program and with the aim of the smooth and open operation of it, an attempt was made to develop a management approach that would preserve both the operation of the program and the monitoring of its progress. For this purpose, after the evaluation of the pilot implementation of the program and before the start of the new school year, an open invitation was made to teachers to take on the role of e-tutor. This process is repeated at the end of each school year where the program is evaluated, the annual report is published and the invitation for new teachers follows. Teachers interested in taking on the role of e-tutor attend a 16-hours training program, which includes 6 hours of face to face and 10 hours of distance training. The program is completed with the educator's micro-teaching to the other educator and to the trainers and the exams in order to achieve his/her certification. Upon obtaining the certification, the educator has the opportunity to take on the role of e-tutor.

With the start of each new school year, preparations are made in order to select new e-tutors from the available team that has been developed. The selections are made under the responsibility of the school directors and for each year 8 mathematics teachers, 8 Greek language teachers and 2 teachers of physics are selected. Then, the program coordinator using a questionnaire invites the 18 new e-tutors to state possible restrictions, in order to structure their program. The next step is the creation of the personal URL address and room for each e-tutor. The coordinator informs the Information Resources Management department, which undertakes to create the virtual rooms and give access to the e-tutors with specific rules and policies. The coordinator then prepares the education of the new students of the school, that is the students of 7th grade. The training of the students lasts two hours and focuses a) on the use of tools, b) on the possibilities offered to students via the e-tutoring program and c) on the proper use of the digital environment. Upon completion of the students' education, the e-tutoring program begins, which lasts until the end of the school year. The program as already mentioned operates daily except Fridays and weekends [17].

The coordinator informs the e-tutors for the material repository. The repository has a specific structure, so that it is easy to use and is constantly enriched with new material, mainly by the e-tutors. In addition, the coordinator gives the e-tutors their URLs and specifies that they have to inform the students about the e-tutoring sessions in two ways. One of these ways is is with sending daily emails with the details of the session, the hours of the session, the URL and the other one is through the Blackboard Learn and the course that has been created for the e-tutoring. During and after each session, the e-tutors record the names and surnames of the students who joined, their classroom, the time they stayed connected to e-tutoring and the reason why they joined e-tutoring. For this purpose, a file has been created in the repository that is completed online by the e-tutors. Moreover, at the end of each session, the e-tutor records the URL of the session that took place, in a file in the repository, since all sessions are recorded and stored for a whole year, for further use and analysis. Access to the above data is open to the coordinator, e-tutors, directors and is offered through collaborative documents.

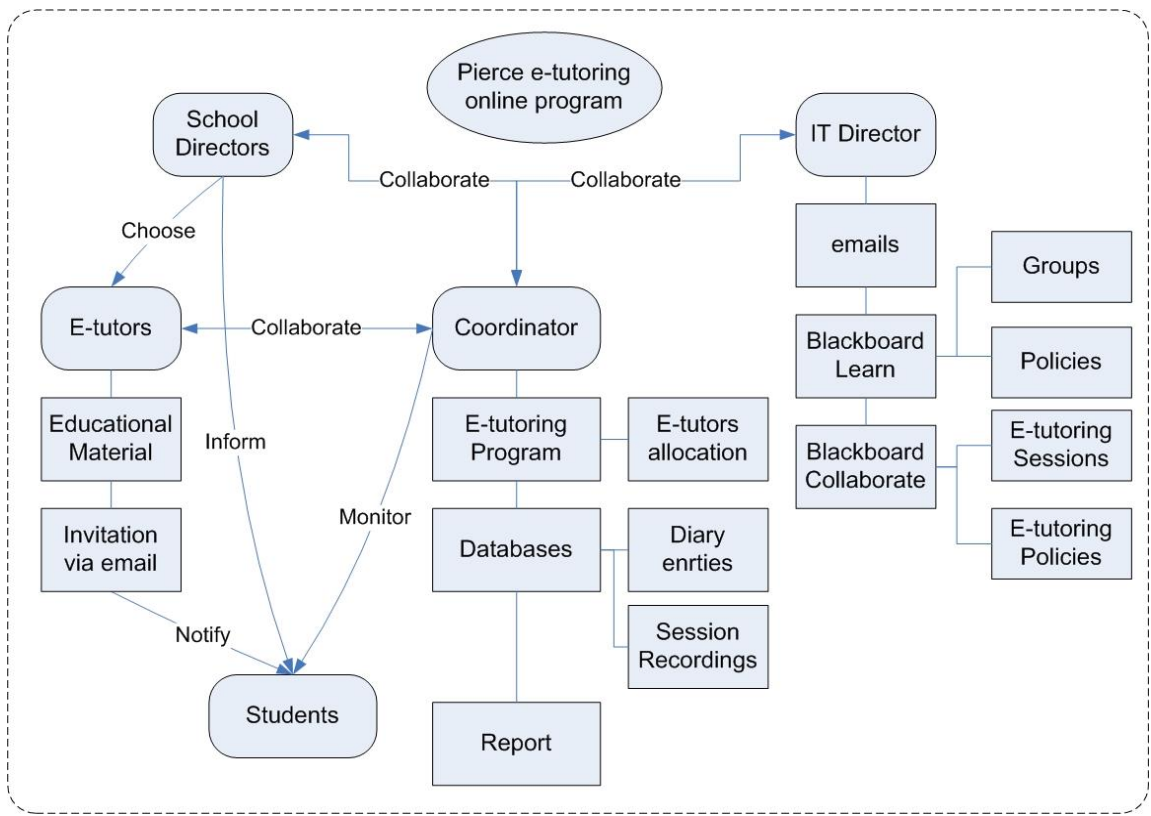

Figure 6. E-tutoring management approach

According to the above data, in the previous school year 2019-2020, 806 connections were made by students of all grades (7th to 10th grade). Of these, 273 were connections for the subject of the Greek language, 412 were connections related to mathematics and 121 were made in physics. Most students come from 7th grade (334 connections) and the fewest from 10th grade (127 connections). The period of the lockdown (March 11, 2020 to May 11, 2020) had a negative effect on e-tutoring sessions, since the participation of students in online courses with the emergency remote teaching, limited the students' need and their willingness to further utilize e-tutoring. The instruction given to the educators for the online learning, foresaw that during the emergency remote teaching the repetition of the material that has already been taught will be done and no new material will be taught. Thus, on the day before the lockdown, the total connections were 730 and during the two months (during which the emergency remote teaching lasted) the connections reached 806.

Twice a year, in order to better coordinate the program, a report is produced with the statistics from the data collected, in order to show the progress of the program and to highlight possible modifications. The structured management framework that has been developed allows the complete depiction of both the processes and the course. Finally, in the second report, suggestions for improvement and further development are made. The whole management approach is presented in Figure 6.

## 5. DISCUSSION

E-tutoring is a dynamic, enjoyable learning perspective that facilitates participation and offers a supportive / complementary way of learning. In addition, it offers opportunities for students and

28

educators to come together and at the same time helps students overcome phobias and difficulties they may have had in the conventional classroom. The operation of e-tutoring programs is an innovative educational approach. E-tutoring programs is a product innovation as they are introduced in education as something new and significantly challenging in relation to the characteristics of other student support structures. In addition, it is a process innovation since it appeared to be a new and significantly improved way of learning. The operation of the program led to the need to have an organizational innovation that offers an open framework of operation and organization. Finally, they are a marketing innovation since their integration was an important intervention that contributed to the school's reputation.

Despite the above particularly innovative features, some issues arise that need further study in the management of e-tutoring programs. Initially, it would be useful for the system to monitor both the academic progress of the participants and their development. Data provided by Korea highlight the significant differences between children who regularly participate in e-tutoring programs and those who do not. However, the positive impact that e-tutoring programs can have on the academic achievements of secondary school students appears to students who live in urban areas [18].

Another issue that emerges is the well-defined preparation of future e-tutors in order to be able to separate the way they operate in the conventional classroom from the way that is needed to work in e-tutoring environments. At this point, investigation is required whether the constant and daily change of the e-tutor can create problems in the quality of learning or if it is difficult for the student to transition from one learning approach to another. In addition, because the ultimate goal of these educational programs is the motivation students to learn and continue to learn throughout their life, the transition from the transactional to the transformational phase is crucial [19].

Despite the short-term learning benefits that students can have (they did better on an exam, wrote better on a test, solved an exercise faster, felt safer in the classroom) and although are particularly important for the learner to feel confidence, what is required is his change of attitude towards learning. To this end, opportunities for more accurate data and comprehensive evaluation can be achieved through data mining and data analysis. Big data make possible to extract information about student performance and learning approaches. Thus, e-tutoring systems could enhance learning if they combined with a variety of tools and integrated into a range of student activities, such as how much time they spend studying and how they use digital resources. In this way, trends, resources that are most effective and problem areas can be identified and therefore up-to-date decisions can be made. In this context, these approaches require a balance between students' privacy and open access to their data for research and educational purposes. For this reason, e-tutoring data is open and accessible so that students, teachers, and parents can benefit from advances in research and analysis. The differentiated approach of each student and the adaptation of the education to the needs of him/her can be enhanced with the e-tutoring programs.

Finally, it seems that there is a lot of room for expansion of e-tutoring in schools. Although research shows that participation decreases with age, educational policy makers need to re-evaluate how these programs are offered by comparing them to other, less cost-effective, and less transformative options. In conclusion, e-tutoring programs and especially the Pierce e-tutoring online program, is a useful service that offered to students. To this end, research on the subject needs to be further strengthened to explore issues related to: a) the integration of Learning Management Systems that will function as the unique system of asynchronous and synchronous learning within the school unit, b) approaches that will enhance student participation, c) the exploration of issues that students seek support for, in order to incorporate other methods of student support, such as videos, artifacts, activities.

## 6. LIMITATIONS

The management model presented in the previous sections is an integrated approach to the implementation of e-tutoring programs using a platform, which has specific features and capabilities. More specifically, although Blackboard Collaborate is a reliable and functional platform for online learning, the interaction it offers needs to be compared with other available platforms. In addition, another limitation in the case of Pierce etutoring online program, is that Blackboard Collaborate constitutes an additional application outside the Learning Management System (LMS) of the school. Finally, the personal data issues that the organization needs to manage may have a negative effect to increase the participation and to offer personal support to specific students.

## 7. CONCLUSIONS

The integration of e-tutoring programs in the educational practice across the countries proves to be possible, feasible and, perhaps, necessary, as the benefits from this integration are multiple. The new reality with the pandemic has significantly changed the image of distance education and it is necessary to take advantage of its positive elements. Thus, as households acquired more technological tools, e-tutoring could work in a way that would meet the need for remedial teaching or additional teaching support. In this context, the possibility of utilizing e-tutoring programs in school education, for all students and schools through existing structures of the Ministries of Education is possible. Bearing in mind the reduced cost of the program and therefore the savings for families and the fact that many households have access to the internet it seems possible to include such programs. In general, e-tutoring programs contribute to the creation of a model of education which, through the alternative way of approaching and supporting learning, manages to meet the needs of students and their families directly and effectively.

**AUTHORS**

**Spyridon Doukakis** earned his PhD in Education & Digital Technologies from the University of the Aegean and he completed his postdoctoral research at Ionian University in the field of Neuroeducation. He is a researcher at Ionian University and educator of Mathematics and Computer Science at Pierce-The American College of Greece. He has worked as a Special Counselor for Teacher Training at the Institute of Educational Policy for the Hellenic Ministry of Education. He has co-authored more than 90 papers in international journals and conferences as well as book chapters. He has received a Fulbright scholarship and has been awarded with the Harvard Prize Book Teacher Award.

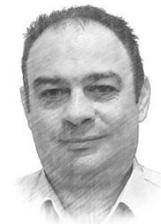